\documentclass[pre,superscriptaddress,twocolumn,amsmath,amssymb,floatfix]{revtex4-1}

\usepackage{graphicx}% Include figure files

\begin{document}

\title{Analytically Solvable Model of Spreading Dynamics with Non-Poissonian Processes} % Force line breaks with \\
%\title{Exact Evaluation of the Effect of Non-Poissonian Processes on Spreading Dynamics} % Force line breaks with \\
\author{Hang-Hyun Jo}
\email{hang-hyun.jo@aalto.fi}
\affiliation{BECS, Aalto University School of Science, Espoo, Finland}
\author{Juan I. Perotti}
\affiliation{BECS, Aalto University School of Science, Espoo, Finland}
\author{Kimmo Kaski}
\affiliation{BECS, Aalto University School of Science, Espoo, Finland}
\author{J\'anos Kert\'esz}
\affiliation{Center for Network Science, Central European University, Budapest, Hungary}
\affiliation{BECS, Aalto University School of Science, Espoo, Finland}

\date{\today}% It is always \today, today,
             %  but any date may be explicitly specified

\begin{abstract}
Non-Poissonian bursty processes are ubiquitous in natural and social phenomena, yet little is known about their effects on the large-scale spreading dynamics. In order to characterize these effects we devise an analytically solvable model of Susceptible-Infected (SI) spreading dynamics in infinite systems for arbitrary inter-event time distributions and for the whole time range. Our model is stationary from the beginning, and the role of lower bound of inter-event times is explicitly considered. The exact solution shows that for early and intermediate times the burstiness accelerates the spreading as compared to a Poisson-like process with the same mean and same lower bound of inter-event times. Such behavior is opposite for late time dynamics in finite systems, where the power-law distribution of inter-event times results in a slower and algebraic convergence to fully infected state in contrast to the exponential decay of the Poisson-like process. We also provide an intuitive argument for the exponent characterizing algebraic convergence.

\end{abstract}


\maketitle

\section{Introduction}

Events of the dynamical processes of various complex systems are often not distributed homogeneously in time but have intermittent or bursty character. This is ubiquitously witnessed in processes of nature like earthquake statistics~\cite{Utsu1995}, solar flare~\cite{Wheatland1998}, and firing of neurons~\cite{Kemuriyama2010}, but also in social processes like financial interactions~\cite{Lillo2003} and human communication activities~\cite{Barabasi2005}. In these examples the distribution of inter-event times follows a power-law behavior~\cite{Utsu1995,Wheatland1998,Kemuriyama2010,Lillo2003,Barabasi2005} in contrast to the homogeneous Poissonian processes showing exponential distribution.

Dynamical processes of complex systems can be considered to take place on a network formed by pairwise interactions between the constituents of the system~\cite{Albert2002,Newman2010}. In the recently developed approach of temporal networks~\cite{Holme2012} a link between two nodes is considered existing only at the moment of interaction. One of the most interesting dynamical processes on networks is spreading~\cite{Newman2005,Castellano2010,Goltsev2012,Perra2012,Boguna2013} that takes place on temporal networks and the statistics of events strongly influences its most important feature, namely the speed of propagation. This feature is of pivotal interest and importance, e.g., for halting epidemic outbreaks or promoting diffusion of innovations. 

Recently much effort has been devoted to clarify how burstiness of events influences the spreading speed, partly by using empirical data analysis~\cite{Vazquez2007,Karsai2011, Iribarren2011, Miritello2011,Rocha2011,Gauvin2013} and partly by model calculations~\cite{Holme2012,Vazquez2007,Iribarren2009,Rocha2013,VanMieghem2013}. In those studies the bursty character of an event sequence was found to slow down the late time dynamics of spreading, evidenced also by a heavy tail in the inter-event time distribution. However, for the early time dynamics, conflicting results have been reported~\cite{Masuda2013b}. In studies by Vazquez~\textit{et al.}~\cite{Vazquez2007} and Karsai~\textit{et al.}~\cite{Karsai2011} the burstiness is found to slow down spreading, while other works point towards the opposite direction~\cite{Iribarren2011,Rocha2011,Rocha2013}. This calls attention to the importance of small inter-event times or to the role of lower bound of inter-event times. The effect of lower bound has been largely ignored, although it is present in empirical phenomena and it is important for understanding the early time behavior of reference systems.

Model studies usually aim to reproduce some empirical observations for to uncover the main mechanisms of the real underlying processes. Here we take the perspective to construct a tractable and analytically solvable model, where the effects of different parts of the inter-event time distribution can be studied explicitly and understood in detail. The model we consider is without correlations found in realistic datasets, except for the correlation due to the inter-event time distribution. In this way we hope to provide a reference system, which can serve as a starting point for later studies.

\section{Model} 

\begin{figure}[!t]
    \includegraphics[width=.85\columnwidth]{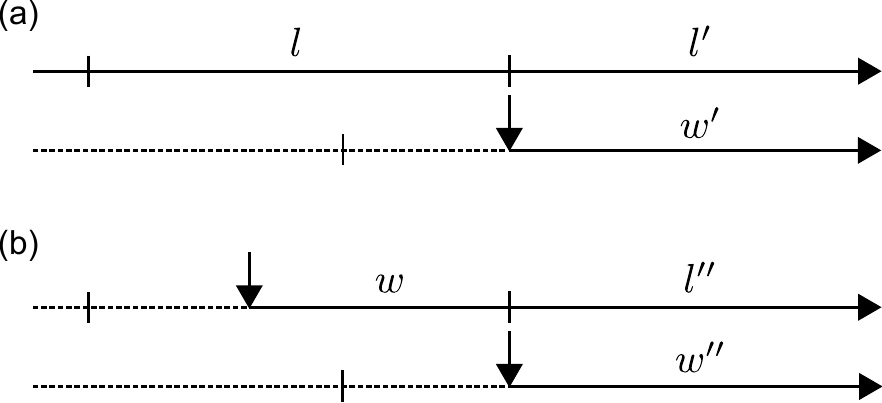}
    \caption{Schematic diagram of the infections by an already infected node (a) and by a newly infected node (b). Vertical lines and vertical arrows denote activation timings of nodes and infections from infected nodes (solid horizontal line) to susceptible nodes (dotted horizontal line). Inter-event times $l$, $l'$, and $l''$ are independent of each other, and so are residual waiting times $w$, $w'$, and $w''$.}
    \label{fig:branching}
\end{figure}

In order to model bursty spreading phenomenon we study deterministic Susceptible-Infected (SI) dynamics taking place on a temporal network of infinite size. Each node remains inactive for an inter-event time, denoted by $l$, before becoming instantaneously active, and then it turns inactive for another inter-event time period, and so on. The inter-event time distribution $P_0(l)$ is assumed to be the same for all nodes, implying a homogeneous population. The activation pattern of a node is independent of whether it is susceptible (S) or infected (I). Whenever any infected node becomes active, it chooses a random node and infects it if the chosen node is susceptible, see Fig.~\ref{fig:branching}. Here the probability of choosing susceptible node is $1$ in the infinite size system as the dynamics starts from a single infected node. The newly infected inactive node should wait a residual waiting time, denoted by $w$, before it becomes active again. The distribution of $w$ is derived from $P_0(l)$ as $P_1(w)=\frac{1}{\mu}\int_w^\infty P_0(l)dl$, with $\mu$ denoting the mean of $l$. Thus the dynamics is stationary from the beginning as it is independent of the nodes being initially active or inactive. Otherwise the early stage of spreading dynamics could be sensitive to the variation of the initial distribution of active or inactive nodes. 

In our model the dynamics can be interpreted to occur on a temporal network in the sense that any pairwise interaction between nodes, defining a link, is instantaneous and annealed. Such links can be interpreted as directed as the inter-event time distribution is considered only for outgoing events of infecting nodes. The spreading dynamics on this temporal network can be related to a class of Bellman-Harris branching processes~\cite{Harris2002,Iribarren2009,Iribarren2011}. It should be noted that temporal inhomogeneities have been considered in a model study by Perra~\textit{et al.}~\cite{Perra2012}, although they took a different approach from ours by using inhomogeneous activities of nodes.

We investigate the spreading dynamics starting from one infected and \textit{active} node at time $t=0$. The number of infected nodes at later time $t$, denoted by $I_0(t)$, remains as $1$ for the inter-event time $l$ given to the initially infected node. After the first infection at $t=l$, $I_0(t)$ can be written as the sum of two numbers: One is for the infecting node and its subsequent infected nodes, which can be denoted by an independent and identical copy of $I_0$ but starting at $t=l$, thus as $I'_0(t-l)$. The other is for the newly infected node and its subsequent infected nodes, similarly denoted by $I'_1(t-l)$. Here $I'_1$ is an independent and identical copy of $I_1$, i.e., the number of infected nodes starting from one infected and \textit{inactive} node. Thus we get
\begin{eqnarray}
  I_0(t)&=&\left\{\begin{tabular}{ll}
    $1$ & \textrm{if}\ $t<l$,\\
    $I_0'(t-l)+I_1'(t-l)$ & \textrm{if}\ $t\geq l$,
  \end{tabular}\right.
\end{eqnarray}
as depicted in Fig.~\ref{fig:branching}(a). Since the newly infected node must wait a residual waiting time $w$ as in Fig.~\ref{fig:branching}(b), the number of infected nodes starting from one infected and inactive node can be written as
\begin{eqnarray}
  I_1(t)&=&\left\{\begin{tabular}{ll}
    $1$ & \textrm{if}\ $t<w$,\\
    $I_0''(t-w)+I_1''(t-w)$ & \textrm{if}\ $t\geq w$,
  \end{tabular}\right.
\end{eqnarray}
where $I''$s are independent and identical copies of $I$. The generating function for $I_0(t)$ is defined as $F_0(z,t)=\sum_{k\geq 0}\Pr[I_0(t)=k]z^k$, and we get
\begin{equation}
  F_0(z,t)= \left\{\begin{tabular}{ll}
    $z$ & \textrm{if}\ $t<l$,\\
    $F_0(z,t-l)F_1(z,t-l)$ & \textrm{if}\ $t\geq l$.
  \end{tabular}\right.
\end{equation}
Here $F_1(z,t)$ is the generating function defined for $I_1(t)$. By taking the expectation over $l$ with $P_0(l)$, one obtains
\begin{equation}
  F_0(z,t)=z\int_t^\infty P_0(l)dl+\int_0^t F_0(z,t-l)F_1(z,t-l)P_0(l)dl.
\end{equation}
Then, the average number of $I_0(t)$ is calculated as
\begin{eqnarray}
  n_0(t)&\equiv& \langle I_0(t)\rangle =\left.\frac{\partial F_0(z,t)}{\partial z}\right|_{z=1}\\
  &=&\int_t^\infty P_0(l)dl+\int_0^t [n_0(t-l)+n_1(t-l)]P_0(l)dl,\nonumber\\
\end{eqnarray}
where $n_1(t)\equiv \langle I_1(t)\rangle$. Taking the Laplace transform gives
\begin{eqnarray}
  \tilde n_0(s)&=& \frac{1-\tilde P_0(s)}{s}+[\tilde n_0(s)+\tilde n_1(s)]\tilde P_0(s),\\
  \tilde n_1(s)&=& \frac{1-\tilde P_1(s)}{s}+[\tilde n_0(s)+\tilde n_1(s)]\tilde P_1(s),
\end{eqnarray}
which straightforwardly leads to
\begin{equation}
  \tilde n_0(s)=\frac{1}{s}+\frac{\tilde P_0(s)}{(s-\mu^{-1})[1-\tilde P_0(s)]},
\end{equation}
where we have used the relation $\tilde P_1(s)=\frac{1}{\mu s}[1-\tilde P_0(s)]$. Then, $n_0(t)$ can be calculated by taking the inverse Laplace transform of $\tilde n_0(s)$ analytically or numerically if necessary. Note that this solution has been obtained for arbitrary inter-event time distributions and for the whole time range, enabling us to evaluate the effect of burstiness on spreading at any stage of dynamics. In contrast to this case of infinite system size the late time behavior of finite systems cannot be investigated analytically.

As for the non-Poissonian bursty processes, they are often characterized by broad inter-event time distributions, such as Gamma and log-normal distributions~\cite{Iribarren2011} and power-law distribution with exponential cutoff~\cite{Vazquez2007,Rocha2013}. Since these distributions have zero lower bound of inter-event times, the effect of lower bound on the early stage of spreading dynamics has then been ignored, despite the importance of the finite lower bound in empirical phenomena. In order to investigate systematically the effect of lower bound as well as the heavy tails of inter-event times, we consider the \textit{shifted} power-law distribution with exponential cutoff:
\begin{equation}
  \label{eq:shiftedNonPoisson_P0}
  P_0(l)= \frac{l_c^{\alpha-1}}{\Gamma(1-\alpha,\frac{l_0}{l_c})} l^{-\alpha}e^{-l/l_c}\theta(l-l_0),
\end{equation}
where $\Gamma$ is the upper incomplete Gamma function, and $\theta$ is the Heaviside step function. $l_0$ and $l_c$ denote the lower bound and the exponential cutoff of $l$, respectively. In case with $l_c\to\infty$, the value of power-law exponent $\alpha$ should be larger than $2$ to guarantee finite $\mu$, i.e., the mean of $l$. The mean is related to other parameters as follows
\begin{equation}
  \label{eq:xyalpha}
  x=y\frac{\Gamma(1-\alpha,y)}{\Gamma(2-\alpha,y)},
\end{equation}
where $x\equiv \frac{l_0}{\mu}$ and $y\equiv \frac{l_0}{l_c}$. Here $0\leq x\leq 1$ because the mean cannot be smaller than the lower bound. When $y=0$, the relation reduces to $x=\frac{\alpha-2}{\alpha-1}$. Note that setting the power-law exponent $\alpha=0$ reduces the distribution to the shifted Poissonian case. 

\subsection{Poissonian processes} 

As the simplest case, the Poissonian process with $P_0(l)= \mu^{-1} e^{-l/\mu}$ results in the solution
\begin{equation}
  \label{eq:Poisson}
  n_0(t)=e^{t/\mu}. 
\end{equation}
Generally, we consider the shifted Poissonian process by setting $\alpha=0$ in Eq.~(\ref{eq:shiftedNonPoisson_P0}), leading to 
\begin{equation}
  \label{eq:shiftedPoisson_P0}
  P_0^{\rm P}(l)=\frac{1}{\mu-l_0}\exp\left(-\frac{l-l_0}{\mu-l_0}\right)\theta(l-l_0).
\end{equation}
Here we have used $l_c=\mu-l_0$ by Eq.~(\ref{eq:xyalpha}). Then, we get 
\begin{equation}
  \label{eq:shiftedPoisson_ns}
  \tilde n_0(s)=\frac{1}{s}+\frac{1}{s-\mu^{-1}} \frac{1}{[(\mu-l_0)s+1]e^{sl_0}-1}.
\end{equation}
For the early time dynamics, by assuming that $s\gg 1$, we obtain
\begin{equation}
  \tilde n_0(s)\approx \frac{1}{s}+\frac{1}{s-\mu^{-1}} \frac{e^{-sl_0}}{(\mu-l_0)s+1},
\end{equation}
which results in the following solution
\begin{eqnarray}
  n_0(t)&\approx& 1+\frac{1}{2-\frac{l_0}{\mu}} \left[ e^{\frac{t-l_0}{\mu}} -e^{-\frac{t-l_0}{\mu-l_0}}\right]\theta(t-l_0)\\
  &\approx& 1+\frac{t-l_0}{\mu-l_0} \theta(t-l_0).
\end{eqnarray}
As for the last line, the exponential functions for $t\ll \mu$ have been expanded. The lower bound $l_0$ delays the first branching while at time later than $t=l_0$ the spreading is speeded up. Let us define a dimensionless spreading rate at the moment of the first branching as follows 
\begin{eqnarray}
  C_0&\equiv& \mu \left.\frac{dn_0}{dt}\right|_{t=l_0^+},\\
  \label{eq:shiftedPoisson_C0}
  C_0^{\rm P}(x)&=&\frac{1}{1-x}.
\end{eqnarray}

Next, we study the late time dynamics, where the late time for infinite size systems corresponds to the intermediate time for finite size systems. Since it is evident that $n_0(t)\sim e^{t/\mu}$ for large $t$, we characterize the asymptotic behavior by defining the coefficient of the leading exponential term as
\begin{eqnarray}
  \label{eq:Cinfty}
  C_\infty&\equiv& \lim_{t\to\infty} n_0(t)e^{-t/\mu}\\
  \label{eq:Cinfty_s}
  &=&\lim_{s\to 0} s\tilde n_0(s+\mu^{-1}).
\end{eqnarray}
Here we have used the final value theorem~\cite{Ogata2010}
\begin{equation}
  \lim_{t\to\infty}f(t)=\lim_{s\to 0} s\tilde f(s)
\end{equation}
with $f(t)=n_0(t)e^{-t/\mu}$. By plugging Eq.~(\ref{eq:shiftedPoisson_ns}) into Eq.~(\ref{eq:Cinfty_s}), we obtain 
\begin{eqnarray}
  \label{eq:shiftedPoisson_Cinfty}
  C_\infty^{\rm P}(x)=\frac{1}{(2-x)e^x+1}.
\end{eqnarray}
This result implies that the finite lower bound suppresses the late time spreading dynamics.

\begin{figure}[!t]
  \includegraphics[width=.9\columnwidth]{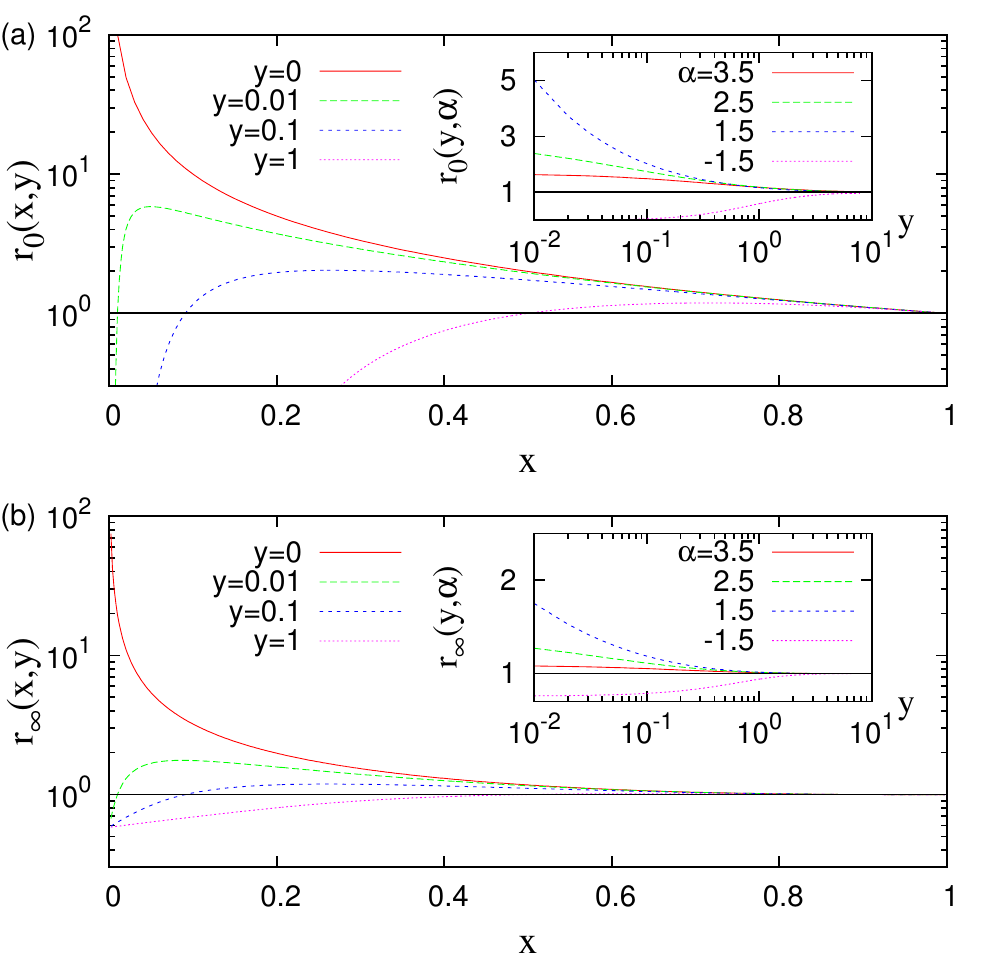}
  \caption{Exact solutions of (a) the ratio of initial spreading rates $r_0(x,y,\alpha)$ in Eq.~(\ref{eq:r0}) and (b) the ratio of asymptotic coefficients $r_\infty(x,y,\alpha)$ in Eq.~(\ref{eq:rInfty}). We note that $x=\frac{l_0}{\mu}$ and $y=\frac{l_0}{l_c}$ with the lower bound $l_0$, the mean $\mu$, and the cutoff $l_c$ of the inter-event time distribution in Eq.~(\ref{eq:shiftedNonPoisson_P0}). We plot $r_0(x,y)$ and $r_\infty(x,y)$ for various values of $y$, and in the insets $r_0(y,\alpha)$ and $r_\infty(y,\alpha)$ for various values of $\alpha$. By definition, the ratios have the value of $1$ for $\alpha=0$. In the limits of $x\to 1$ and/or $y\to \infty$, one gets $P_0(l)\to\delta(l-\mu)$, thus $r_0$ and $r_\infty$ have the value of $1$.}
    \label{fig:ratio}
\end{figure}

\subsection{Non-Poissonian processes} 

Now we consider the general form of $P_0(l)$ in Eq.~(\ref{eq:shiftedNonPoisson_P0}). The Laplace transform of $P_0(l)$ is obtained as follows
\begin{equation}
  \label{eq:shiftedNonPoisson_P0s}
  \tilde P_0(s)=(sl_c+1)^{\alpha-1}\frac{\Gamma(1-\alpha,y(sl_c+1))}{\Gamma(1-\alpha,y)}.
\end{equation}
To investigate the early time dynamics of $n_0(t)$, we consider the case with $s\gg 1$. By expanding the incomplete gamma function, we obtain
\begin{eqnarray}
  \tilde P_0(s)&\approx& \frac{y^{-\alpha}e^{-y}}{\Gamma(1-\alpha,y)}\frac{e^{-sl_0}}{sl_c+1}\\
  \tilde n_0(s)&\approx& \frac{1}{s}+A\left(\frac{1}{s-\mu^{-1}}-\frac{1}{s+l_c^{-1}}\right)e^{-sl_0}
\end{eqnarray}
with $A=\frac{1}{x+y}\frac{y^{1-\alpha}e^{-y}}{\Gamma(1-\alpha,y)}$, leading to 
\begin{eqnarray}
  n_0(t)&\approx& 1+A\left(e^{\frac{t-l_0}{\mu}}-e^{-\frac{t-l_0}{l_c}}\right)\theta(t-l_0)
\end{eqnarray}
The spreading rate at $t=l_0^+$, i.e., $C_0$ is obtained as
\begin{equation}
  \label{eq:shiftedNonPoisson_C0}
  C_0(x,y,\alpha)=\frac{1}{x}\frac{y^{1-\alpha}e^{-y}}{\Gamma(1-\alpha,y)}.
\end{equation}
Note that $x$, $y$, and $\alpha$ are not independent by means of Eq.~(\ref{eq:xyalpha}), and that $C_0(x,y,0)=\frac{1}{1-x}=C_0^{\rm P}(x)$. For the comparison to shifted Poissonian processes, we define the ratio of spreading rates as
\begin{equation}
  \label{eq:r0}
  r_0(x,y,\alpha)\equiv \frac{C_0(x,y,\alpha)}{C_0^{\rm P}(x)},
\end{equation}
which turns out to be exactly the same as the ratio of 
\begin{equation}
  \label{eq:r0PP}
  \frac{P_0(l=l_0^+)}{P_0^{\rm P}(l=l_0^+)}.
\end{equation}
This indicates that the probability of having $l=l_0^+$ determines the early time spreading dynamics. In addition to $r_0(x,y,0)=1$ by definition, it is found that $r_0>1$ for $\alpha>0$ and $r_0<1$ for $\alpha<0$, see Fig.~\ref{fig:ratio}(a). Provided that $\alpha>0$, one can conclude that the non-Poissonian bursty activity always accelerates the early time spreading dynamics as compared to the shifted Poissonian case with the same mean $\mu$ and the same lower bound $l_0$. 

\begin{figure}[!t]
  \includegraphics[width=.9\columnwidth]{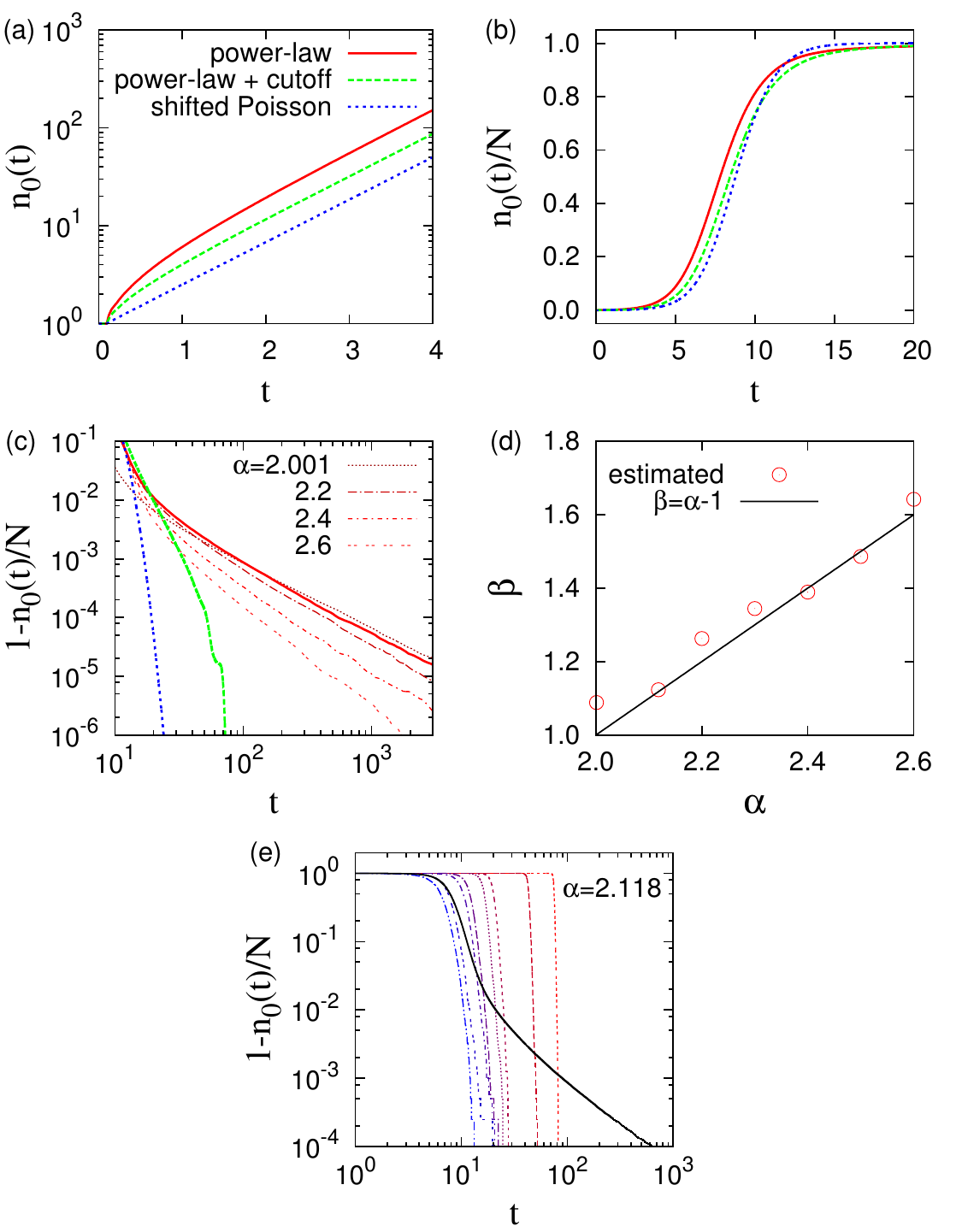}
  \caption{Numerical simulation results of Poisson-like and non-Poissonian cases for systems with infinite size (a) and finite size $N=4\cdot 10^3$ (b-e). In (a-b), we set $\mu=1$ and $l_0\approx 0.106$ for all cases, from which we get $\alpha\approx 2.118$ for $l_c\to\infty$ (power-law) and $\alpha=1.5$ for $l_c=100 l_0$ (power-law with cutoff). Each curve is averaged over up to $1.5\cdot 10^6$ runs. In (c), we plot the fractions of susceptible nodes, $1-\frac{n_0(t)}{N}$, using the same results in (b), as well as for various values of $\alpha$ in the power-law case with $\mu=1$ (thin reddish curves). The Poisson-like case is fitted with $e^{-t}$, while the power-law cases are fitted with $t^{-\beta}$. The estimated $\beta$ as a function of $\alpha$ is presented in (d), comparable with $\beta=\alpha-1$. For understanding this scaling relation, in (e) we plot the individual runs with different initial inter-event times (from blue to red curves) with their average (black solid curve) in case with $\alpha\approx 2.118$.}
    \label{fig:numerics}
\end{figure}

For the late time dynamics, we focus on the asymptotic behavior characterized by $C_\infty$ in Eq.~(\ref{eq:Cinfty}). Similarly to the Poissonian case, one gets the following general result:
\begin{equation}
  C_\infty(x,y,\alpha)=\frac{ (1+\frac{x}{y})^{\alpha-1}\Gamma(1-\alpha,x+y) }{\Gamma(1-\alpha,y)-(1+\frac{x}{y})^{\alpha-1}\Gamma(1-\alpha,x+y)}.
\end{equation}
Note that $C_\infty(x,y,0)=\frac{1}{(2-x)e^x-1}=C_\infty^{\rm P}(x)$. Similarly to $r_0$, we define the ratio
\begin{equation}
  \label{eq:rInfty}
  r_\infty(x,y,\alpha)\equiv \frac{C_\infty(x,y,\alpha)}{C^{\rm P}_\infty(x)}.
\end{equation}
In addition to $r_\infty(x,y,0)=1$ by definition, it is found that $r_\infty>1$ for $\alpha>0$ and $r_\infty<1$ for $\alpha<0$, see Fig.~\ref{fig:ratio}(b). This implies that the non-Poissonian bursty activity accelerates the late time dynamics as compared to the corresponding Poissonian processes. Our analysis is confirmed by the numerical simulations as depicted in Fig.~\ref{fig:numerics}(a).

The correction term to the exponential growth for late time dynamics is obtained for the case of $y=0$ such that 
\begin{equation}
  \label{eq:n0t_correct}
  n_0(t)\simeq C_\infty e^{t/\mu}(1-Be^{-\lambda t}).
\end{equation}
Both $B$ and $\lambda$ can be analytically obtained by taking the Laplace transform of the above equation for $s>\mu^{-1}$. By defining $\epsilon\equiv s-\mu^{-1}>0$, Eq.~(\ref{eq:n0t_correct}) is transformed to 
\begin{equation}
  \epsilon \tilde n_0(\mu^{-1}+\epsilon)=C_\infty\left(1-\frac{B\epsilon}{\lambda+\epsilon}\right).
\end{equation}
By expanding both sides up to the order of $\epsilon^2$ for small $\epsilon$ and comparing the coefficients, one can get $\lambda$ and $B$ as functions of $x$ and $\alpha$.
In the limit of $x\to 0$ ($\alpha\to 2$), we obtain $\lambda\to \mu^{-1}$ and $B\to 1$, resulting in 
\begin{equation}
  n_0(t)\simeq C_\infty e^{t/\mu}-C_\infty.
\end{equation}
The constant term implies the existence of residual waiting times that are effectively infinite due to the non-normalizability of $P_1(w)$, whose tail is characterized by the exponent $\alpha-1$.

\subsection{Finite size effects} 

Finally, we consider the effect of finite system size $N$ on the spreading dynamics. Whenever an infected node becomes active at time $t$, it chooses a random node and infects it if the chosen node is susceptible. In other words, the infection occurs with probability $\frac{N-n_0(t)}{N-1}\approx 1-\frac{n_0(t)}{N}$, but otherwise it does not occur. Since the exact solution could not be obtained, we perform numerical simulations to obtain the spreading dynamics shown in Fig.~\ref{fig:numerics}(b-e). The early and intermediate time dynamics are consistent with the early and late time dynamics of the infinite system, respectively. For the late time dynamics, the non-Poissonian bursty activity results in a slower and algebraic convergence to the fully infected state, i.e., $\sim t^{-\beta}$ with $\beta=\alpha-1$ for the power-law case, in contrast to the exponential decay of the Poisson-like process, i.e., $\sim e^{-t}$.

We provide an intuitive argument for the relation $\beta=\alpha-1$ in the case of power-law inter-event time distributions. While the average fraction of susceptible nodes decays algebraically, the fraction of susceptible nodes for each run turns out to stay almost $1$ and then to suddenly decay exponentially, shown in Fig.~\ref{fig:numerics}(e). The period of staying almost $1$ must be governed mostly by the inter-event time initially given to the first infected node. Therefore, the average fraction of susceptible nodes can be obtained as the fraction of runs that did not reach the fully infected state at time $t$. Such fraction of runs is equal to the probability of having $l>t$, which is proportional to $t^{-(\alpha-1)}$, leading to $\beta=\alpha-1$. Considering the dominant role of inter-event time given to the first infected node, this argument should be valid for the SI dynamics with power-law inter-event time distribution on a broad class of networks.

\section{Conclusions}

We have introduced an analytically solvable model for studying the effect of non-Poissonian bursty inter-event time distributions on the Susceptible-Infected (SI) spreading dynamics. Our model is devised to be stationary from the beginning. For this, we make a realistic assumption that for each infection event the infecting node should wait another inter-event time and the newly infected node should wait a residual waiting time before them becoming active again. With this assumption, we could obtain the analytic solution of spreading dynamics in infinite systems for arbitrary inter-event time distributions but more importantly for the whole time range. By our analytic solution, the role of lower bound of inter-event times has been exactly compared for Poisson-like and non-Poissonian processes. We also note that, as done in case of null models, randomizing or shuffling the event timings to destroy temporal correlations can eliminate the lower bound of inter-event times. Hence for systematic comparison between the original situation and the null model one needs to employ a shuffling method that conserves the lower bound.

Let us next discuss apparently conflicting results for the early stage spreading dynamics presented in~\cite{Karsai2011,Rocha2011}. The early stage dynamics is mainly driven by small inter-event times, which generally leads to the faster spreading for non-Poissonian cases than for Poisson-like cases. This is the case only when any infected node can always find a susceptible node without a topological limit. However, it is well known that the mobile phone call network (MCN) in~\cite{Karsai2011} has the community structure accompanying the bottleneck effect due to weak links between communities~\cite{Onnela2007}. The large inter-event times associated with such weak links are the main reason for slowing down of spreading on the MCN, while the role of small inter-event times coupled with local topological structure is still important for spreading within communities. This provokes us to study more realistic models as a future work. On the other hand, the spreading is enhanced by the burstiness on the sexual network~\cite{Rocha2011}. This might be because the sexual network has different community and/or temporal structures from the MCN, so that the spreading on the sexual network can be better understood by our model to some extent.

As a follow-up our model can be extended to incorporate a number of other complex situations, such as Susceptible-Infected-Recovered (SIR) spreading dynamics and cascading phenomena. Our results should be of interest beyond the community of network scientists because non-Poissonian processes are ubiquitous and yet little is known about their impact on the large-scale dynamics.

\begin{acknowledgments}
Financial supports from Aalto University postdoctoral programme (HJ), from the Academy of Finland, project No. 260427 (JIP), from the Academy of Finland's Center of Excellence programme 2006-2011, project No. 129670 (KK), and from the TEKES FiDiPro and FP7 MULTIPLEX projects (JK) are gratefully acknowledged.
\end{acknowledgments}

%\bibliography{h2jo-spreading}
%merlin.mbs apsrev4-1.bst 2010-07-25 4.21a (PWD, AO, DPC) hacked
%Control: key (0)
%Control: author (0) dotless jnrlst
%Control: editor formatted (1) identically to author
%Control: production of article title (0) allowed
%Control: page (1) range
%Control: year (0) verbatim
%Control: production of eprint (0) enabled
%

\end{document}